\documentclass[a4paper,11pt]{article}
\usepackage{pos}
\usepackage{bbding}
\usepackage{slashed}
\usepackage{subfigure}

\newcommand{\mc} {\mathcal}

\title{Seesaw Models of Neutrino Masses on the Convex Cone of Positivity Bounds}

\author*[a,b]{Xu Li}
\author[a,b]{Shun Zhou}

\affiliation[a]{Institute of High Energy Physics, Chinese Academy of Sciences,\\
 Beijing 100049, China}

\affiliation[b]{School of Physical Sciences, University of Chinese Academy of Sciences, \\
 Beijing 100049, China}

\emailAdd{lixu96@ihep.ac.cn}
\emailAdd{zhoush@ihep.ac.cn}

\abstract{The convex geometric framework of positivity bounds allows us to explore the ultraviolet (UV) states in new physics models from the bottom up. The UV states in three types of seesaw models for tiny Majorana neutrino masses, as irreducible representations of the ${\rm SU}(2)^{}_{\rm L}$ gauge group, naturally fit into this framework. Since neutrino masses arising from the dimension-five Weinberg operator imply that the UV states may couple to the left-handed lepton doublet $L$ and the Higgs doublet $H $, we construct a convex cone of positivity bounds to constrain the dimension-eight operators consisting of $L$ or $H$. Such a construction offers a novel way to distinguish between different seesaw models and derive lower bounds on the seesaw scale.}

\FullConference{%
  41st International Conference on High Energy physics - ICHEP2022\\
  6-13 July, 2022\\
  Bologna, Italy
}

\begin{document}
\maketitle

\section{Introduction}
The origin of neutrino masses is one of the tantalizing puzzles of Standard Model (SM), while it can be explained in the framework of the Standard Model effective field theory (SMEFT) through the dimension-five (dim-5) Weinberg operator ${\cal O}^{(5)} \equiv \overline{L} \tilde{H} \tilde{H}^{\rm T} L^{\rm c}$ \cite{Weinberg:1979sa}. Therefore, it is very likely that new physics accounting for neutrino masses is connected to the left-handed lepton doublet $L$ and the Higgs doublet $H$. The type-I, type-II and type-III seesaw models are tree-level ultraviolet (UV) completions of the Weinberg operator, but they are indistinguishable up to dim-5. To explore the origin of neutrino masses, one may turn to higher-dimensional operators in the corresponding low-energy effective field theories (EFT's).

In this connection, the positivity bounds, which are robust constraints on the Wilson coefficients (WC's) in the EFT arising from the axiomatic principles of quantum field theories, prove to be very helpful. Based on the geometrical perspective ~\cite{Zhang:2020jyn}, the positivity bounds that apply to dimension-eight (dim-8) operators, can be used to reconstruct the UV theory from low-energy observables, known as the ``inverse problem''. In this formalism, the positivity bounds are normal vectors of a convex cone in the dim-8 WC space. The UV states living in the irreducible representations (irrep's) of  the symmetries of the $S$-matrix are projected into the WC space and form a convex hull. The heavy particles in the seesaw models are all in the irrep's of the SM gauge symmetry, implying a profound relation between the origin of neutrino masses and the EFT with dim-8 operators. Solving the inverse problem provides us with a novel way to identify the seesaw model and understand the nature of neutrino masses. 

\section{Theoretical Formalism}
We focus on the second derivative of the forward 2-to-2 amplitudes ${\cal M}^{}_{ij \to kl}(s,t \to 0)$ with respect to $s$ (where $s,t $ are the ordinary Mandelstam variables, and the indices $i,j,k,l$ are the particle indexes), which will extract the $s^2$ dependence of amplitude and we define it as
\begin{flalign}
	\begin{aligned}
		M^{ijkl} \equiv \lim_{s \to 0} \frac{d^2\mathcal{M}^{}_{ij\rightarrow k l}
			\left(s\right)}{ds^2} \; .
		\label{eq:tensor1}
	\end{aligned}
\end{flalign}
Utilizing the analyticity and unitarity, as well as the generalized optical theorem at the tree level, the dispersion relation can recast Eq.~(\ref{eq:tensor1}) into~\cite{Bi:2019phv}
\begin{equation}
	\begin{aligned}
		M^{ijkl}&
		=\frac{1}{2\pi}\int_{(\varepsilon\Lambda)^2}^{\infty}{\frac{d\mu}{\mu^3}\sum_{X \in \mathbf{r}}\left[\mathcal{M}_{ij\rightarrow X}
			\mathcal{M}_{kl\rightarrow X}^\ast+\left(j\leftrightarrow l\right)\right]} \; ,
		\label{eq:tensor2}
	\end{aligned}
\end{equation}
where the summation is over all the intermediate UV states $X$ and $(j \leftrightarrow l)$ is the crossing channel. We choose $X$ to be the irrep $\mathbf{r}$ of the ${\rm SU}(3) \otimes {\rm SU}(2)^{}_{\rm L} \otimes {\rm U}(1)^{}_{\rm Y}$ gauge group, the particles $i$ and $j$ belong to the irrep $\mathbf{r}^{}_i$ and $\mathbf{r}^{}_j$, respectively. By the decomposition rule $\mathbf{r}^{}_i \otimes \mathbf{r}^{}_j = \sum_\alpha C_{\mathbf{r},\alpha}^{i,j} \mathbf{r}$, where $C_{\mathbf{r},\alpha}^{i,j}$ are the Clebsch-Gordan (CG) coefficients and the summation over all the states $\alpha$'s in $\mathbf{r}$ is implied, we can rewrite Eq.~(\ref{eq:tensor2}) as below:
\begin{flalign}
	M^{ijkl}=\frac{1}{2\pi}\int_{\left(\varepsilon\Lambda\right)^2}^{\infty}{\frac{d\mu}{\mu^3}\sum_{X\text{\ in\ }\mathbf{r}}
		{\left|\left\langle X\middle|\mc{M}|\mathbf{r}\right\rangle\right|^2 \mc{G}_\mathbf{r}^{ijkl}}} \; ,
	\label{eq:projector}
\end{flalign}
with $\mc{G}_\mathbf{r}^{ijkl}\equiv \sum_\alpha C^{i,j}_{\mathbf{r},\alpha}
\left(C^{k,l}_{\mathbf{r},\alpha}\right)^\ast + (j \leftrightarrow l)$ 
being defined as the ``generator", and $M^{ijkl}$ can be generated by positive combinations of $\mc{G}_\mathbf{r}^{ijkl}$'s from different irrep's, which means $M^{ijkl}$ is constrained inside a convex cone.
The extremal ray (ER) of the convex cone is defined as the element that cannot be decomposed into any positive sum of other elements in the cone, and the generator may plays the role of an ER, the cone is defined by $\mathcal{C} = \text{cone}\left\{\mathcal{G}_{\mathbf{r}}^{ijkl}\right\}$. One observation from Eq.~(\ref{eq:tensor2}) is that the UV state residing in the irrep of the symmetry group corresponds to the ER of the cone. 

Now the primary goal is to examine the positions of seesaw models in the convex cone, and to identify them with the help of dim-8 operators. We notice the UV states of heavy particles in three types of seesaw models are in the irrep's of $\mathbf{1},\mathbf{3},\mathbf{3}$ of ${\rm SU}(2)^{}_{\rm L}$ gauge group, respectively, so they naturally fit into this framework. Given all known symmetries in the theory, it is straightforward to find all UV states that lead to dim-8 operators. 

\section{The UV States}
To produce the Weinberg operator, the new physics should be coupled to lepton and Higgs doublets at the same time. Therefore, we examine the minimal space of WC's, that is, exploring the positivity bounds on the amplitude involving $H$ or $L$. The dim-8 operators that can contribute to such process are classified into three types in Table~\ref{tab:Opera}, in which the $\sigma^I$ (for $I = 1, 2, 3$) stand for the Pauli matrices and $\overleftrightarrow{D_\nu} \equiv D^{}_\mu - \overleftarrow{D^{}_\mu}$ with $D^{}_\mu$ being the covariant derivative in the SM has been defined. For simplicity,  we will consider only one lepton flavor.
\begin{table}
	\scalebox{0.85}{
	\renewcommand\arraystretch{0.9}
	\setlength{\tabcolsep}{2 mm}
	\begin{tabular}{ c | c | c }
		\hline \hline 
		$LLHH$ & $LLLL $& $HHHH$  \\
		\hline \hline
		${\cal O}^{}_1= (\bar{L}\gamma_\mu {\rm i} \overleftrightarrow{D_\nu} L ) \left(D^\mu H^\dag  D^\nu H\right)$ & ${\cal O}^{}_3 = \partial_\nu\left(\bar{L}\gamma^\mu L\right)\partial^\nu\left(\bar{L}\gamma_\mu L\right)$ & ${\cal O}^{}_5=\left(D_\mu H^\dag D_\nu H\right)\left(D^\nu H^\dag D^\mu H\right)$\\
		${\cal O}^{}_2= (\bar{L}\gamma_\mu\sigma^I {\rm i} \overleftrightarrow{D_\nu} L) \left(D^\mu H^\dag\sigma^I D^\nu H\right) $ & ${\cal O}^{}_4 = \partial_\nu\left(\bar{L}\gamma^\mu\sigma^I L\right)\partial^\nu\left(\bar{L}\gamma_\mu\sigma^I L\right)$ & ${\cal O}^{}_6=\left(D_\mu H^\dag D_\nu H\right)\left(D^\mu H^\dag D^\nu H\right)$\\
		 & & ${\cal O}^{}_7=\left(D_\mu H^\dag D^\mu H\right)\left(D_\nu H^\dag D^\nu H\right)$ \\
		\hline \hline
	\end{tabular}
	}
	\vspace{0.01 cm}
	\caption{Independent operators consisting of the lepton doublet $L$ or the Higgs doublet $H$ at dimension-eight. Note that there are also two  $LLHH$ type dim-8 operators, but they don't contribute to the amplitudes in the forward limit.}
	\label{tab:Opera}
\end{table} 

The positivity bounds on the $HHHH$-type operators and $LLLL$-type operators are studied in the Ref.~\cite{Remmen:2019cyz} and Ref.~\cite{Fuks:2020ujk}, respectively. In the present paper, we enlarge the space of WC's by further combining those two $LLHH$-type operators ${\cal O}^{}_1$ and ${\cal O}^{}_2$ in Table~\ref{tab:Opera}. 
The key to construct a convex cone is to use the CG coefficients to obtain the generator matrix in Eq.~(\ref{eq:projector}). We use the fact that both $L$ and $H$ belong to irrep $\mathbf{2}$ of the ${\rm SU}(2)^{}_{\rm L}$, then list all the CG coefficients given by the decomposition $ \mathbf{2}\times \mathbf{2}= \mathbf{1}+\mathbf{3}$ and 
$ \mathbf{2}\times \bar{\mathbf{2}}= \mathbf{1}+\mathbf{3}$ as follows:

\begin{table}
	\centering 
	\scalebox{0.85}{
	\renewcommand\arraystretch{0.3}
	\setlength{\tabcolsep}{2 mm}
	\begin{tabular}{ c c c c c c c }
		\hline \hline State & Spin& Charge & Interaction & Seesaw &  ER & $\vec{c}$ \\
		\hline \hline$E$ & $1/2$ & ${\bf 1}_{-1}$ & $g \bar{E}\left(H^{\dagger} L\right)$ & & \CheckmarkBold &$\displaystyle \frac{1}{2}(-1,-1,0,0,0,0,0)$\\
		$\Sigma_1$ & $1/2$&$ {\bf 3}_{-1}$ & $g \bar{\Sigma}^I_1 \left(H^{\dagger} \sigma^{I} L\right)$ 
		& & \XSolidBrush &$\displaystyle \frac{1}{2}(-3,1,0,0,0,0,0)$\\
		$N$ & $1/2$&$ {\bf 1}_{0}$ & $g \bar{N}\left(H^{\rm T} \epsilon L\right)$ &$\operatorname{Type-I}$ & \CheckmarkBold
		&$\displaystyle \frac{1}{2}(-1,1,0,0,0,0,0)$ \\
		$\Sigma$ & $1/2$&$ {\bf 3}_{0}$ & $g \bar{\Sigma}^I \left(H^{\rm T} \epsilon \sigma^{I} L\right)$ 
		& $\operatorname{Type-III}$ & \XSolidBrush & $\displaystyle \frac{1}{2}\left(-3,-1,0,0,0,0,0\right)$\\
		$\mathcal{B}_1$ & $1$&${\bf 1}_{1}$ & $g \mathcal{B}_1^{\mu} \left[(H^{\dag} \epsilon {\rm i} \overleftrightarrow{D_{\mu}} H^\ast)
		+ \frac{x}{M} (\bar{L^{\rm c}} \epsilon {\rm i} \overleftrightarrow{D_{\mu}} L)\right]$ 
		& &  \XSolidBrush &$\displaystyle \frac{1}{2}(0,0,x^2,-x^2,16,0,-16)$\\
		$\Xi_1$ & $0$&${\bf 3}_{1}$ & $g \Xi^{I}_1 \left[M (H^{\dag} \epsilon \sigma^{I} H^\ast )+ x (\bar{L^{\rm c}} 
		\epsilon \sigma^{I} L )\right]$ & $\operatorname{Type-II}$& \XSolidBrush & $\displaystyle \frac{1}{2}\left(0,0,-3 x^{2}, - x^{2},0,16,0\right)$ \\
		$\mathcal{S}$ & $0$ & ${\bf 1}_{0 \rm S}$ & $g M \mathcal{S}\left(H^{\dagger} H\right)$ & & \CheckmarkBold  &$2(0,0,0,0,0,0,1)$\\
		$\mathcal{B}$ & $1$ & ${\bf 1}_{0 \rm A}$ & $g \mathcal{B}^{\mu}\left[H^{\dagger}{{\rm i}\overleftrightarrow{D_{\nu}}} 
		H+x (\bar{L} \gamma_{\mu} L)\right]$ & &\XSolidBrush  &$\displaystyle \frac{1}{2}\left(0,0,-x^{2},0, -4,4,0\right)$ \\
		$\Xi$ & $0$ & ${\bf 3}_{0 \rm S}$ & $g M \Xi^{I} (H^{\dagger} \sigma^{I} H )$ & & \XSolidBrush & $2(0,0,0,0,2,0,-1)$\\
		$\mathcal{W}$ & $1$ & ${\bf 3}_{0 \rm A}$ & $g \mathcal{W}^{I \mu}\left[ (H^{\dagger} \sigma^{I} 
		{\rm i}\overleftrightarrow{D_{\mu}} H )
		+ x (\bar{L} \gamma_{\mu} \sigma^{I} L)\right]$ & &\XSolidBrush & $\displaystyle \frac{1}{2}\left(0,0,0,-x^{2},4,4,-8\right)$\\
		\hline\hline
	\end{tabular}
	}
	\vspace{0.01 cm}
	\caption{All possible UV states in irreps that can be coupled to the lepton doublet $L$ and the Higgs doublet $H$, where the spin, the change of the ${\rm SU}(2)^{}_{\rm L} \otimes {\rm U}(1)^{}_{\rm Y}$ group [the subscripts ``S" and ``A" refer to symmetric and antisymmetric representations] and the corresponding vectors in WC space are shown, where $x$ is an arbitrary real number. }
	\label{tab:UVstates}
\end{table}
\begin{equation}
	\begin{aligned}
		&C_{\mathbf{1},c}^{ab} = \epsilon^{ab} \; , \quad C_{\mathbf{3},c}^{ab} = \left(\epsilon \sigma^I\right)^{ab} \;, {\bar{C}}_{\mathbf{1},c}^{ab} = \delta_b^a \; , \quad \, {\bar{C}}_{\mathbf{3},c}^{ab} = \left(\sigma^I\right)^a_b \; ,
	\end{aligned}
\end{equation}
where $\epsilon \equiv {\rm i}\sigma^2$ and the subscript ``$c$" is trivial for ${\bf 1}$ but $c = I$ for ${\bf 3}$.  
The obtained generators can be matched into the WC space, i.e. $\mathcal{G}^{ijkl}_{\mathbf{r}} = \sum_i c_{\mathbf{r}}^i M^{ijkl}_i$ for irrep ${\mathbf{r}}$, which can be effectively labelled as a vector $\vec{c}_{\mathbf{r}}$. Physically, the existence of generator $\mathcal{G}^{ijkl}_{\mathbf{r}}$ in WC space corresponds to a heavy state living in irrep ${\mathbf{r}}$, and appearing as a propagator in the $s$ channel of process $ij\to kl$. From this point of view, we can write down all UV theories for the tree-level process of the $L$ and $H $ scattering, and summarize them in Table~\ref{tab:UVstates}. For each heavy state, it's spin, the quantum number ${\bf r}^{}_{Y}$ under the ${\rm SU}(2)^{}_{\rm L}\otimes {\rm U}(1)^{}_{\rm Y}$ group, and it's interactions with SM particles are given. Each UV state corresponds a unique vector $\vec{c}$ in the WC space. As a result, the convex hull of these vectors forms the positivity cone. Besides, the ERs of the cone are also indicated in one column. In particular, we mark the seesaw models of neutrino masses, which naturally appear in the cone.

\begin{figure}[t]
	\centering  
	\subfigure[3D cross section of the 5D $LLLL+HHHH$ subspace]{
	\label{fig:3Dspace}
	\includegraphics[width=0.5\linewidth]{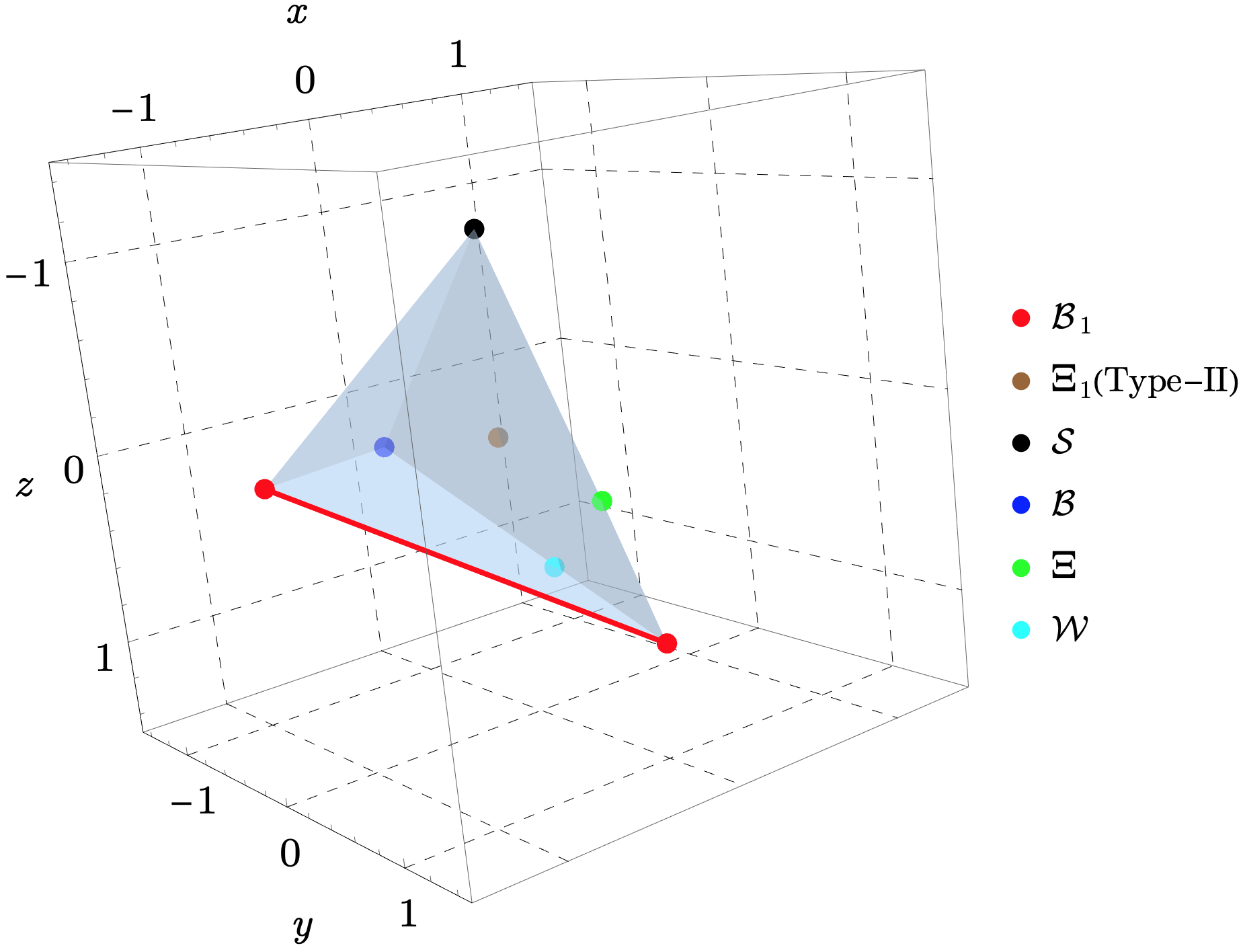}}
	$\ \ \ $
	\subfigure[Convex cone in the $LLHH$ subspace]{
	\label{fig:2Dspace}
	\includegraphics[width=0.4\linewidth]{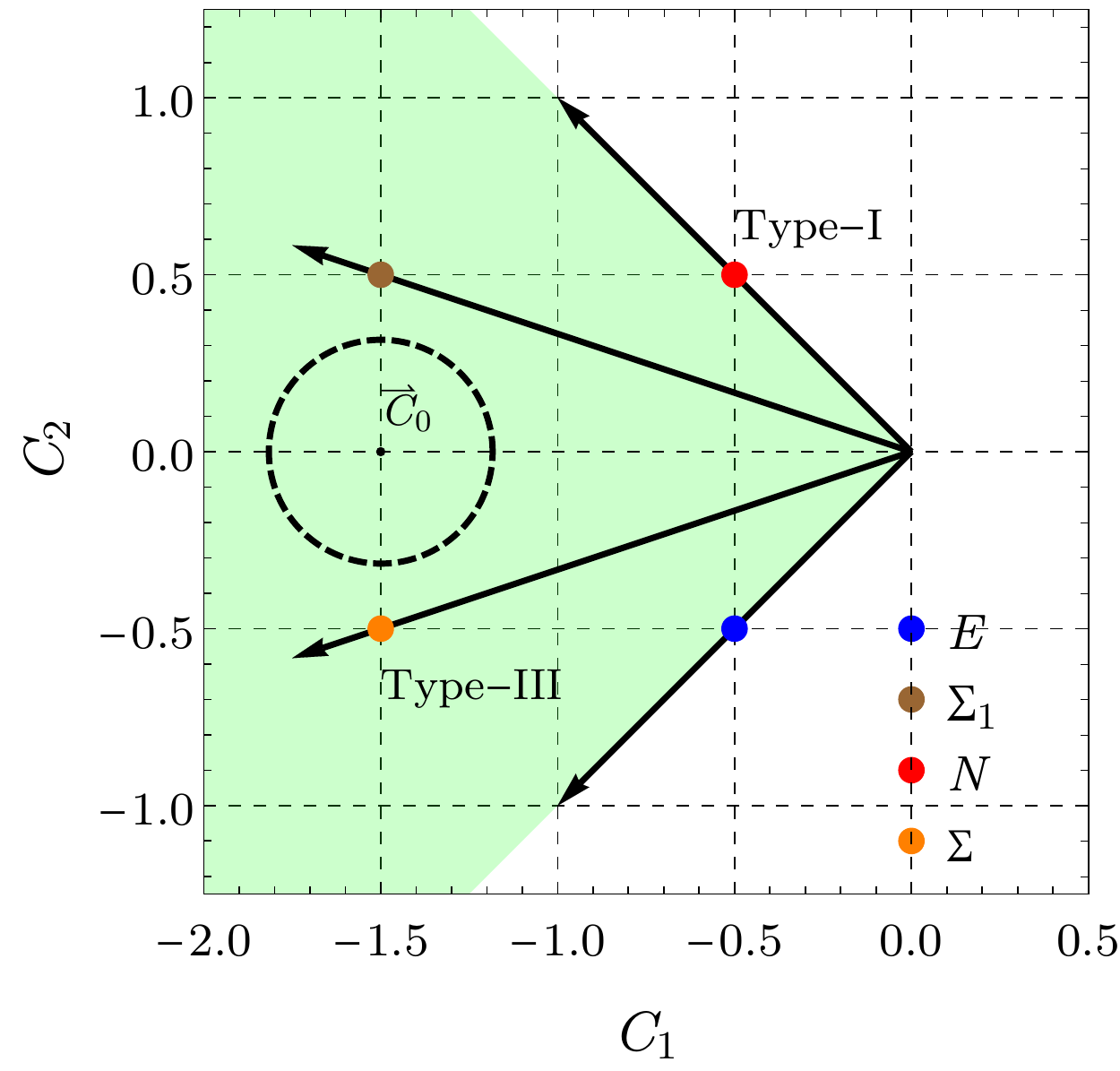}} 	
	\caption{The geometric structure of the convex cone, where different scenarios of UV completion are represented by colored dots or lines. 
	In Fig.~\ref{fig:3Dspace}, to show the 5D space in a 3D cross section, we have chosen a particular projection direction.
	Fig.~\ref{fig:2Dspace} is a 2D space, only the green region is allowed by positivity bounds.
	}
	\label{fig:convexcone}
\end{figure}

For the forward scattering, a boson is not able to be the tree-level completion of the $LLHH$-type operators, while the fermion cannot mediate the 4-$H$ or 4-$L$ scattering. According to this observation, the $ LLHH$ subspace should be separated from $HHHH$ and $LLLL$, as shown in Table~\ref{tab:UVstates}. This observation allows us to discuss the $LLHH$ subspace without worrying about the other WC's in the next section. The $LLLL+HHHH$ subspace is a 5D space, see Fig.~\ref{fig:3Dspace}. However, if we restrict ourselves into the $HHHH$ subspace corresponding to last three components of $\vec{c}$, positivity bounds, $C_6\geq 0,C_5+C_6\geq 0,C_5+C_6+C_7\geq 0$ can be derived, The classical results in Ref.~\cite{Remmen:2019cyz} are reproduced. Similarly, we can reach the result in Ref.~\cite{Fuks:2020ujk} if we only concern the $LLLL$ space, positivity bounds are: $C_3+C_4\leq 0,C_4\leq 0$. 
The above bounds are intact even in the enlarged space, because the ERs of these two subspaces remain unchanged. In addition, the vector $\vec{c}$'s don't contain a component linear in $x$, causing subspaces of $LLLL$ and $HHHH$ can be discussed separately. However, the convex geometric analysis of $LLHH$ subspace will lead us to new bounds.

\section{The $LLHH$ Subspace}
For the $LLHH$ subspace, we focus on the first two components $c_1$ and $c_2$ of the $\vec{c}$ vector. We draw the convex cone in Fig.~\ref{fig:2Dspace}. Each vector in Table~\ref{tab:UVstates} is marked with a colored point to represent a UV state. The convex hull of these vectors forms a green positivity region. Only the coefficients falling in the region of cone are allowed. The ER's are the two edges of the cone, namely,  $\vec{c}=(-1/2, -1/2)$ and $(-1/2,1/2)$. Extracting the normal vectors of the two edges will provide us the following positivity bounds:
\begin{equation}
C_1 + C_2  \leq 0,\quad  C_1 - C_2  \leq 0 \; .
\end{equation}

One direct application of positivity bound is to solve the inverse problem, i.e. to infer the information about the UV physics. Especially, we notice that the type-I seesaw and type-III seesaw appear in the 2D subspace $LLHH$. The type-I seesaw appears as one of the edges of the cone, and the type-III seesaw is inside the cone. This fact allows us to constrain the scales of the two seesaw models by measuring $C_1$ and $C_2$. For example, one can probe the pair production of the Higgs bosons via $e^+ e^- \to hh$ in future electron-positron colliders.

Suppose the global fit from the experiment provides us a $\Delta \chi^2$ range for coefficients $C_1$ and $C_2$, as shown by the point $\vec{C}_0$ and a circle in Fig.~\ref{fig:2Dspace}. Eq.~(\ref{eq:projector}) tells us that $\vec{C}_0$ can be interpreted as a positive linear combination of irreps, namely, $\vec{C} = \sum_i \omega_i \vec{c}_i$ with $i=E, N, \Sigma, \Sigma_1$. It is important that the $\omega_i$ contains the information of the UV state that we want to know, because $\omega_i = g_i^2/M_i^4$. Interestingly, with the knowledge of convex geometry, we can directly extract the upper bound on $\omega_i$ from $\vec{C}_0$, thereby limit the coupling and mass of possible new physics~\cite{Fuks:2020ujk}.
The upper bound on $\omega_i$ can be derived by finding the maximal value of $\lambda$ such that the following vector breaks the positivity condition
\begin{equation}
\vec{C}(\lambda) \equiv \vec{C}_0-\lambda \vec{c_i} = \sum_{j \neq i} \omega_j \vec{c}_j + (\omega_i-\lambda) \vec{c}_i \; .
\label{eq:optimization}
\end{equation}
The value of $\lambda$ can be stated as the maximum possibility for the UV state $i$ to exist and explain the experimental data. In particularly, due to the convex nature of the WC space, we first point out that Eq.~(\ref{eq:optimization}) can be identified as a conic optimization problem~\cite{Li:2022tcz}. However, the existence of uncertainty may obscure our extraction of upper bounds.

\section{Summary}
We have examined three classes of dim-8 operators involving lepton and Higgs doublets, and revealed the geometric structure of the convex cone of positivity bounds in the subspace of the relevant WC's at the tree level. This framework provides a novel way to distinguish the seesaw models and their analogues. We study the subspace formed by two $LLHH$-type operators in detail. The type-I seesaw model resides on one of edges of the cone, and type-III seesaw model lives inside the convex. 
We also explain how to extract the constraints on the UV theories once the experimental measurements of the WC's of dim-8 operators are available. These ideas are expected to be realized in the future lepton colliders.

\section*{Acknowledgements}

This work was supported in part by the National Natural Science Foundation of China under grant No. 11835013 and the Key Research Program of the Chinese Academy of Sciences under grant No. XDPB15.

\end{document}